\documentclass{article}
\usepackage{graphics}
\title{\bf Eternally accelerating universe without event horizon}
\author{ Pedro F. Gonz\'{a}lez-D\'{\i}az.\\
Instituto de Matem\'{a}ticas y F\'{\i}sica Fundamental\\ Consejo Superior
de Investigaciones Cient\'{\i}ficas\\ Serrano 121, 28006 Madrid,
SPAIN\\ }
\date{October 15, 2001}
\begin{document}
\maketitle \large \setlength{\baselineskip}{0.9cm}

\begin{center}
{\bf Abstract}
\end{center}
Increasing astronomical evidence indicates that the expansion of
the universe is accelerating. By simply solving Einstein equations
we show in this letter that a wide class of generic quintessence
models leading to eternal acceleration is associated with
spacetime static metrics which do not exhibit future event
horizons, and therefore do not pose the same problems for string
theory as asymptotic de Sitter spaces.

\pagebreak

String theory is now considered by most theorists as the
fundamental framework to successfully deal with all small-scale
phenomena, including quantum-gravity effects. However, it has
recently been noted [1,2] that as it comes to the foundations of
cosmology - and particularly modern precision cosmology, string
theory does not appear to prove as successful. The recent growing
astronomical evidence that the expansion of the universe has
recently begun to accelerate [3] is actually posing rather
stringent challenges to string theory. If the today dominating
vacuum dark energy is all absorbed into the form of a cosmological
constant (i.e. if the universe is entering an eternal de Sitter
phase), then it has been noticed that no de Sitter space
backgrounds can be accommodated within the framework of the
controllable mathematics of string theory [4,5]. The reason
essentially is that in de Sitter space there necessarily is a
future event horizon and hence the space will contain regions that
are inaccessible to light probes, which does not allow introducing
the $S$-matrix or $S$-vector description required in string theory
[1,4,5].

In addition, it has been also argued [2,6] that in case that the
accelerating expansion of the universe be originated by a
quintessence, slow-varying field with sufficiently negative
pressure $p=\omega\rho$, if the state-equation parameter $\omega$
is kept constant in future evolution, the universe will inexorably
accelerate forever, eventually showing an event horizon which
again leads to the above-alluded problem for string theory.
However, while the incompatibility of de Sitter space with string
theory appears fully unavoidable, the case for quintessence seems
still to offer some room for a peaceful coexistence between string
theory and cosmology [7,8]. Actually, whereas the existence of an
event horizon has been long proved for de Sitter space, nobody has
hitherto mathematically shown (though it has been generally
assumed) that an eternally accelerating universe driven by a
quintessence field be associated with spacetime metrics that
inexorably show event horizons. The aim of this paper is to
investigate whether an eternally accelerating universe would
necessarily show a future event horizon. This research will be
here carried out by simply obtaining static, spherically symmetric
solutions of the Einstein equations that correspond to the vacuum
dark energy associated with a perfect fluid which is made of a
homogeneous, slowly-varying quintessence field for different
constant state equations. Our final conclusion is that for the
entire range of possible negative pressures (other than
$p=-\rho/3$ and $p=-\rho$) which lead to an accelerating universe,
there exist static spacetime metrics which do not exhibit event
horizons.

We start by introducing a metrical {\it ansatz} with Lorentzian
signature given by
\begin{equation}
ds^2=-A(r)dt^2 +B(r)dr^2 +r^2 d\Omega_2^2 ,
\end{equation}
where $A(r)$ and $B(r)$ are the metric tensor components which
depend only on the radial coordinate $r$, and $d\Omega_2^2$ is the
metric on the unit two-sphere. For a generic state equation of the
quintessence field $p=\omega\rho$ and spatial curvature
$\kappa=+1$, the Einstein equations can be written as [9]
\begin{equation}
8\pi
G\omega\rho=\frac{1}{B}\left(\frac{1}{r^2}+\frac{A'}{rA}\right)
-\frac{1}{r^2}
\end{equation}
\begin{equation}
8\pi G\omega\rho =\frac{1}{4B}\left(\frac{2A''}{A}-
\frac{(A')^2}{A^2}-\frac{2B'}{rB}
-\frac{A'B'}{AB}+\frac{2A'}{rA}\right)
\end{equation}
\begin{equation}
8\pi G\rho=-\frac{1}{B}\left(\frac{1}{r^2}
-\frac{B'}{rB}\right)+\frac{1}{r^2} ,
\end{equation}
in which the prime ($'$) denotes derivative with respect to the
radial coordinate $r$. To these equations we have to add [9] the
gravitational equations for the energy-momentum tensor,
$T^{k}_{i;k}=0$, which in the present case lead to the general
expression for the energy density $\rho$
\begin{equation}
\alpha\rho=A(r)^{-(1+\omega)/(2\omega)},
\end{equation}
with $\alpha$ an arbitrary integration constant. From Eqns. (2)
and (4) we can further derive
\begin{equation}
\left(\frac{1+3\omega}{2\omega}\right)\int\frac{dr(1-
B)}{rB}A^{(1+\omega)/(2\omega)}= \frac{2\pi(\omega-1)}{\alpha}-
\frac{A^{(1+\omega)/(2\omega)}}{B}-K_1 ,
\end{equation}
where $K_1$ is an integration constant.

We solve in what follows the above field equations, restricting to
cases within the interval of $\omega$-values which lead to an
accelerating behaviour for the expanding universe, including the
extreme situations $\omega=-1/3$ and $\omega=-1$. The latter case
describes the upper end point of quintessence models and can be
easily solved (note that all dependence on $A(r)$ in Eqns. (5) and
(6) is ruled out for $\omega=-1$) to produce for $K_1=-1$ the
well-known de Sitter solution
\begin{equation}
A(r)=B(r)^{-1}= 1-\frac{8\pi Gr^2}{3\alpha},
\end{equation}
where one can isolate the cosmological constant to be given by
$\Lambda=8\pi G/\alpha$. Of course, this corresponds to an
eternally accelerating universe which will eventually show an
event (cosmological) horizon at $r=\sqrt{3/\Lambda}$ and,
therefore, keeps the kind of fundamental physics problems which
were alluded previously [1-6]. The other extreme case,
$\omega=-1/3$, corresponds to the onset of the accelerating regime
for a universe without any ordinary (observable+dark) matter
content. It already has been considered for any value of $\kappa$
by Chernin, Santiago and Silbergleit [10] who denoted quintessence
for a constant state equation $\omega=-1/3$ as {\it Einstein
quintessence} because it actually describes an Einstein static
universe [11]. A general solution can be in this case readily
obtained because the integral term in the right-hand-side of Eqn.
(6) vanishes for $\omega=-1/3$. Thus, from the Einstein equations
we obtain
\begin{equation}
\frac{8\pi Gr^2}{3\alpha} +\frac{1}{AB}+K_1=0 ,\;\;\;
B=c_0\left(\frac{A' r^2}{\sqrt{A}}\right)^2 ,
\end{equation}
with $c_0$ again an integration constant, whose general solution
reads
\begin{equation}
A(r)= K_3 -\frac{8\pi GK_2}{3K_1 \alpha}\left(1+\frac{3K_1
\alpha}{8\pi Gr^2}\right)^{1/2}
\end{equation}
\begin{equation}
B(r)=-\frac{1}{\left(K_1+\frac{8\pi Gr^2}{3\alpha}\right)A(r)} ,
\end{equation}
where $K_1$, $K_2$ and $K_3$ are arbitrary constants, unless by
the condition $K_1 K_3=-1$. This solution possesses a singularity
at $r=0$.

In the solution given by the metric tensor components (9) and (10)
one can generally distinguish the following three situations. (i)
If $K_1 <0, K_3
>0$ (for any value of $K_2$), then there will be an apparent
singularity at $r_h=\sqrt{3\alpha|K_1|/(8\pi G)}$. This does not
represent a conventional event horizon since, while the metric for
$r>r_h$ has a Kleinian signature (- - + +), the metric for $r<r_h$
is complex. (ii) If $K_1 >0, K_2 <0$ and $K_3 <0$, there will be a
kind of event horizon at
\[r_h=\left(\frac{3\alpha K_1 K_3^2}{8\pi GK_2^2} -\frac{8\pi
G}{3\alpha K_1}\right)^{-1} ,\] which marks the transition from a
Kleinian metric for $r<r_h$ to an Euclidean (Riemannian) metric
for $r>r_h$. Finally, (iii) when $K_1 , K_2 >0$ and $K_3 <0$ there
is no coordinate singularity (no event horizon) and the metric
keeps an Euclidean signature everywhere.

Of much greater interest for the aim of the present study are the
cases which correspond to an accelerating universe $-1/3
>\omega>-1$ for which cases the deceleration parameter [12]
\begin{equation}
q_0=\frac{1}{2}\left[\Omega_M +\Omega_{\phi}(1+3\omega)\right]
\end{equation}
(where $\Omega_M$ and $\Omega_{\phi}$ are the specific energy
densities for ordinary (observable+dark) matter and the
quintessential field) becomes negative even for moderate $\Omega_M
>0$. Because for this regime none of the terms and metric tensor
components present in Eqn. (6) are canceled, it is very difficult
to obtain a general solution of the Einstein equations. None the
less, starting with the general expressions
\begin{equation}
\frac{A'}{A}+\frac{B'}{B}=\frac{8\pi
G(1+\omega)r}{\alpha}A^{-(1+\omega)/(2\omega)}B
\end{equation}
\begin{equation}
B=1
+\frac{\left[(A')^{-2\omega/(1+\omega)}Ar^2\right]'}{A(A')^{-(3\omega
+1)/(\omega+1)}\left(A' r+\frac{2(3\omega+1)A}{\omega+1}\right)} ,
\end{equation}
which are obtained from the field equations, one can still obtain
some simple solutions for any particular constant values of the
quintessential state equation parameter $\omega$, covering the
entire range of possible accelerating universes, except the
extreme Einstein quintessence ($\omega=-1/3$) and de Sitter
($\omega=-1$) cases. Actually, inserting Eqns. (12) and (13) in
the field equations, we can also derive a general differential
equation for the metric component $A(r)$ for any constant value of
$\omega$ within the accelerating regime $-1/3
>\omega >-1$
\begin{equation}
\left(\frac{A'
A^{-(1+\omega)/(2\omega)}}{r^{(1+2\omega)/\omega}}\right)'
+\left(\frac{A^{(\omega-1)/(2\omega)}}{r^{(1+
3\omega)/\omega}}\right)' +\frac{\alpha\left(A' r^2\right)'}{8\pi
G\omega r^{(1+6\omega)/\omega}}=0 .
\end{equation}

A simple solution for $A(r)$ for any $\omega$ in the accelerating
range $-1<\omega <-1/3$ can now be obtained, and hence using Eqn.
(13) the corresponding solution for $B(r)$ can also be derived.
These metric components can finally be worked out to read:
\begin{equation}
A(r)= Kr^{4\omega/(1+\omega)}
\end{equation}
\begin{equation}
B=\frac{\omega^2 +6\omega +1}{(1+\omega)^2} ,
\end{equation}
where the component $B\equiv B(\omega)$ is a constant which is
negative definite for all $\omega$ in the whole accelerating range
$-1<\omega <-1/3$. The parameter $K$ appearing in Eqn. (15) is
also a constant coefficient which depends on $\omega$ and the
quintessence vacuum energy density as follows:
\begin{equation}
K\equiv K(\omega,\alpha)= \left(\frac{2\pi G(\omega^2
+6\omega+1)}{\omega\alpha}\right)^{2\omega/(1+\omega)} .
\end{equation}
All the solutions given by Eqns. (15) - (17) show: (i) a curvature
singularity at $r=0$, (ii) a Kleinian definite signature, (- - +
+), and (iii) {\it no} event horizon. The latter property appears
to prevent any of the shortcomings that one would expect when
defining the $S$-matrix and/or the proper observables in a theory
described by a finite dimensional Hilbert space if the universe is
accelerating [1-6]. This would circumvent any challenging
questions for string theories, unless we consider the vacuum dark
energy to be originating from a cosmological constant so that the
spacetime is given by the de Sitter solution [4,5]. Clearly, the
price to be paid for this rather comfortable property is to have a
static spacetime metric with a Kleinian signature where the metric
component $g_{rr}$ is constant. Of course, one can always continue
our general solution
\begin{equation}
ds^2 =-K(\omega,\alpha)r^{4\omega/(1+\omega)}dt^2 +
\frac{\left(\omega^2+6\omega+1\right)dr^2}{(1+\omega)^2} +r^2
d\Omega_2^2
\end{equation}
into a metric with Lorentzian signature by introducing
suitable coordinate rotations, such as
\[\sqrt{B}r\rightarrow \bar{r},\;\;
\theta\rightarrow\sqrt{B}\bar{\theta} ,\;\; \phi\rightarrow
-i\sqrt{B}\bar{\phi}, \] to obtain
\[ds^2 =-K(\omega,\alpha)\bar{r}^{4\omega/(1+\omega)}dt^2 +
d\bar{r}^2+\bar{r}^2\left(d\bar{\theta}^2+
\sinh^2(\sqrt{-B}\bar{\theta})d\bar{\phi}^2\right) ,\] which
obviously shows a breakdown of the original spherical symmetry. A
similar breakdown can be obtained using the coordinate change
$\Omega_2\rightarrow\bar{\Omega}_2+\sqrt{-B}\ln r$, which now
converts metric (18) into
\[ds^2=-K(\omega,\alpha)r^{4\omega/(1+\omega)}dt^2
+2\sqrt{-B}rd\bar{\Omega}_2 dr+ r^2d\bar{\Omega}_2^2 .\]

The main result of the present work is the realization that,
contrary to what has been currently believed, cosmological
spacetimes which are accelerating due to the presence of
quintessential vacuum energy, are described by static metrics that
show no event horizon. This result is made possible due to the
Kleinian character of the metric signature. Actually, one may
argue that such spacetimes possess nonchronal regions, as it can
be checked by the coordinate change
\begin{equation}
r={\ell}\tan^2\left(\frac{\psi}{2}\right) ,\;\;\;
{\ell}=K^{-(1+\omega)/(4\omega)},
\end{equation}
with the new coordinate being half-periodic, running from $\psi=0$
($r=0$) to $\psi=\pi$ ($r=\infty$). In terms of this coordinate,
metric (18) can be written as
\begin{equation}
ds^2= -\tan^{8\omega/(1+\omega)}\left(\frac{\psi}{2}\right)dt^2
+B{\ell}^2\frac{\tan^2\left(\frac{\psi}{2}\right)
d\psi^2}{\cos^4\left(\frac{\psi}{2}\right)}
+{\ell}^2\tan^4\left(\frac{\psi}{2}\right)d\Omega_2^2 .
\end{equation}
Let us now consider a world line for matter defined by
$\theta,\phi$ = const., $t=-\beta\phi$, where $\beta$ is a real
constant. In this case, metric (20) reduces to
\[ds^2= -\left[\beta^2
\tan^{8\omega/(1+\omega)}\left(\frac{\psi}{2}\right)
-B{\ell}^2\frac{\tan^2\left(\frac{\psi}{2}\right)
}{\cos^4\left(\frac{\psi}{2}\right)}\right]d\psi^2 .\] Thus, since
$B$ is a negative definite constant in the accelerating range
$-1<\omega<-1/3$ and provided the parameter $\beta$ is real, this
element of line is timelike for such a range. It follows that an
observer moving on the world line will have an increasingly
negative time coordinate and, even though she cannot reach the
point $\psi=\pi$, her evolution will be confined to follow a path
within a half-closed timelike curve. It appears therefore that the
price to be paid for having an accelerating universe without
future event horizon, and hence for having quintessence
cosmological models that can peacefully coexist with string
theory, is to allow for spacetimes with nonchronal regions which
evolves along half-closed timelike curves in such models.

\vspace{.8cm}

\noindent{\bf Acknowledgements} The author thanks David Santiago
of Stanford University for useful comments. This work was
supported by DGICYT under Research Project No. PB97-1218.

\pagebreak

\noindent\section*{References}

\begin{description}
\item [1] W. Fischler, A. Kashani-Poor, R. McNees and S. Paban,
JHEP 0107 (2001) 003.
\item [2] S. Hellerman, N. Kaloper and L. Susskind, JHEP 0106
(2001) 003.
\item [3] S. Perlmutter {\it et al.}, Astrophys. J. 483 (1997)
565; S. Perlmutter {\it et al.} (The Supernova Cosmology Project),
Nature 391 (1998) 51; P.M. Garnavich {\it et al.} Astrophys. J.
Lett. 493 (1998) L53; B.P. Schmidt, Astrophys. J. 507 (1998) 46;
A.G. Riess {\it et al.} Astrophys. J. 116 (1998) 1009.
\item [4] T. Banks, {\it Cosmological breaking of supersymmetry or
little Lambda goes back to the future}, hep-th/0007146; E. Witten,
{\it Quantum gravty in DeSitter space}, hep-th/0106109 .
\item [5] T. Banks and W. Fischler, {\it M-theory observables
for cosmological space-times}, hep-th/0102077 .
\item [6] Xiao-Gang He, {\it Accelerating universe and event
horizon}, astro-ph/0105005; S.M. Carroll, Phys. Rev. Lett. 81
(1998) 3067.
\item [7] J.D. Barrow, R. Bean and J. Mageijo, Mon. Not. R. Astron.
Soc. 316 (2000) L41 .
\item [8] See related works in: S.M. Carroll, {\it Dark energy and
the preposterous universe}, astro-ph/0107571; J.W. Moffat,
astro-ph/0108201; T. Chiuet and Xiao-Gang He, {\it The
accelerating universe: A gravitational explanation},
astro-ph/0107453; E.H. Gudmundsson and G. Bjornsson, {\it Dark
energy and the observable universe}, astro-ph/0105547 (Astrophys.
J., en press); J. Ellis, N.E. Navromatos and D.V. Nanopoulos, {\it
String theory and an accelerating universe}, hep-th/0105206 .
\item [9] L.D. Landau and E.M. Lifshitz, {\it Teor\'{\i}a Cl\'{a}sica de
Campos} (Revert\'{e}, Barcelona, Spain, 1966).
\item [10] A.D. Chernin, D.I. Santiago and A.S. Silbergleit, {\it
Interplay between gravity and quintessence: A set of new GR
solutions}, astro-ph/0106144 .
\item [11] S.W. Hawking and G.F.R. Ellis, {\it The Large Scale
Structure of Space-Time} (Cambridge Univ. Press, Cambridge, UK,
1972).
\item [12]B.P. Schmidt {\it et al.} Astrophys. J. 509 (1998) 54.

\end{description}

\end{document}